\documentclass[]{spie}  
\usepackage[]{graphicx}
\usepackage[]{aas_macros}

\title{Studying accreting black holes and neutron stars with time
  series: beyond the power spectrum} 

\author{Simon Vaughan\supit{a} and Philip Uttley\supit{b}
\skiplinehalf
\supit{a}Department of Physics and Astronomy, University of
Leicester, Leicester, LE1 7RH, UK\\
\supit{b}School of Physics and Astronomy, University of
Southampton, Southampton SO17 1BJ, UK
}

\authorinfo{E-mail: sav2@star.le.ac.uk}

\begin{document} 
  \maketitle 

\begin{abstract}
The fluctuating brightness of cosmic X-ray sources, particularly accreting black holes and neutron star systems, has enabled enormous progress in understanding the physics of turbulent accretion flows, the behaviour of matter on the surfaces of neutron stars and improving the evidence for black holes. Most of this progress has been made by analysing and modelling time series data in terms of their power and cross spectra, as will be discussed in other articles in this volume. Recently, attempts have been made to make use of other aspects of the data, by testing for non-linearity, non-Gaussianity, time asymmetry and by examination of higher order Fourier spectra. These projects, which have been made possible by the vast increase in data quality and quantity over the past decade, are the subject of this article. 
\end{abstract}



\section{INTRODUCTION}
\label{sect:intro}  

\begin{figure}
  \begin{center}
    \begin{tabular}{c}
			\includegraphics[height=5.8cm]{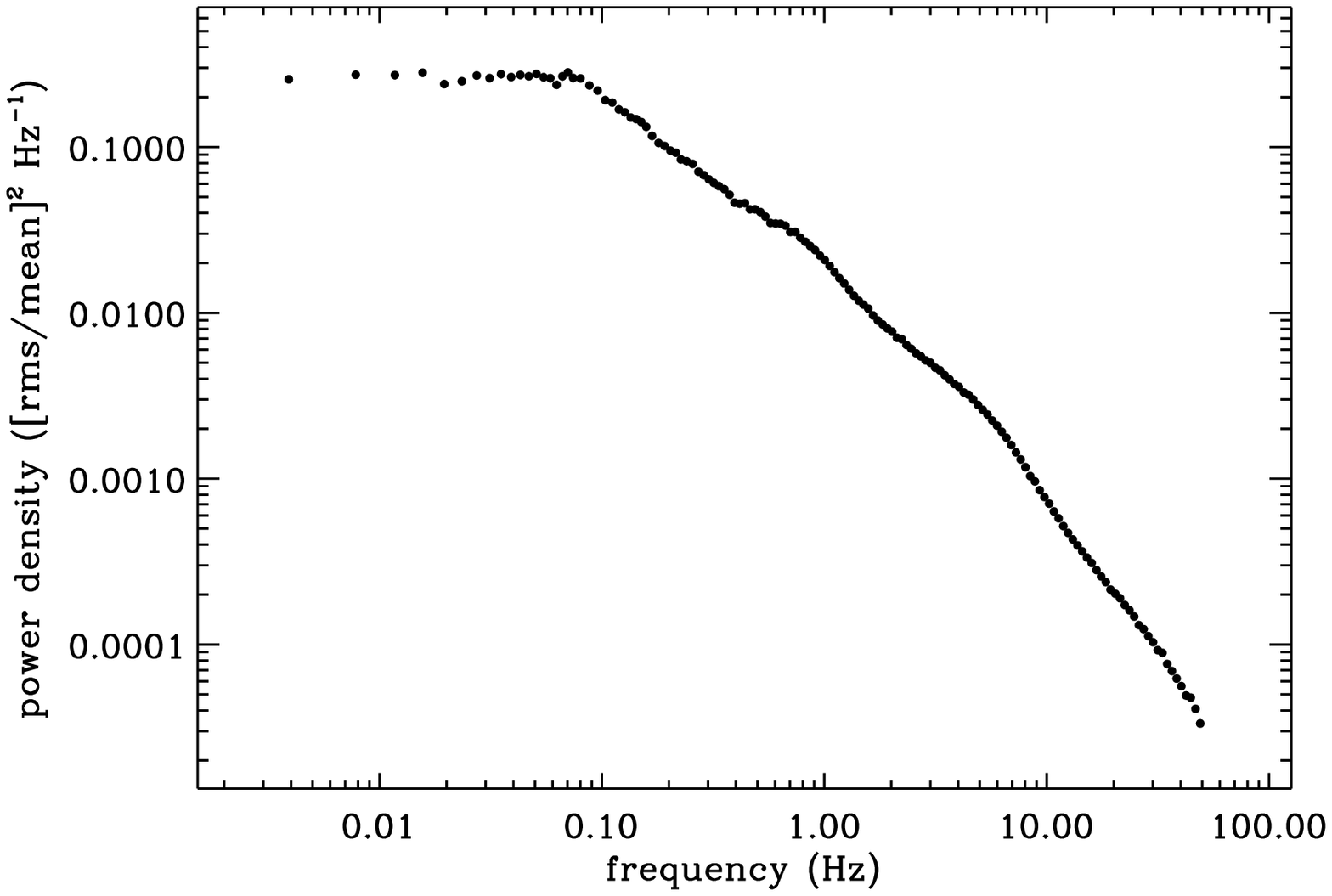}
			\includegraphics[height=5.8cm, angle=0]{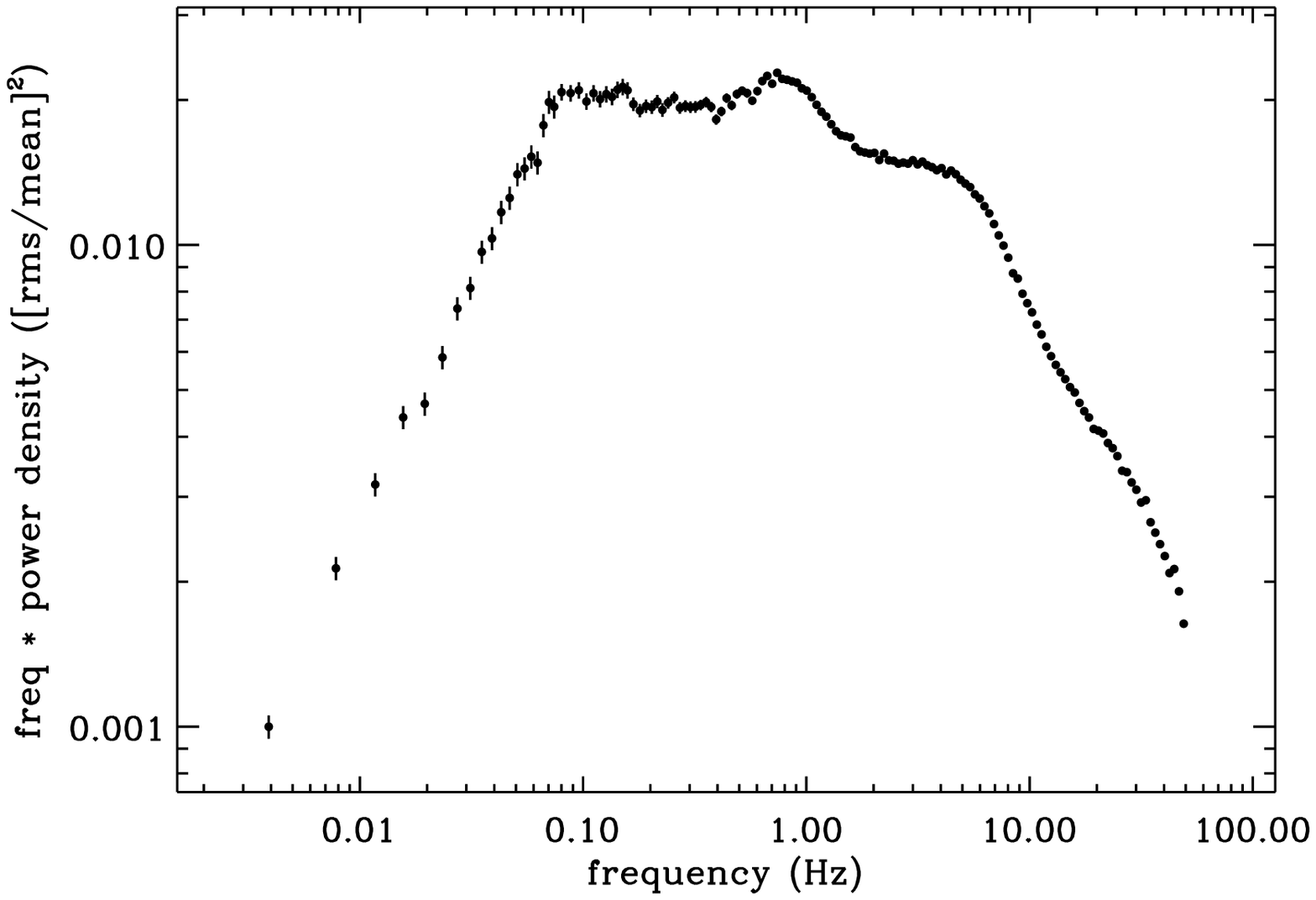}
  	\end{tabular}
  \end{center}
	\caption{\label{fig:psd} 
Power spectrum of the X-ray count rate from an \emph{RXTE} observation of the accreting black hole Cygnus X-1 taken on 16-19 Dec 1996, sampled in bins of $\Delta t = 2^{-8}$ s ($3.90625$ ms).
On the left is shown the power density, revealing the approximately broken power law spectral shape, with a $1/f^{0}$ spectrum below $0.06$~Hz, a roughly $1/f$ spectrum in the range $\sim 0.06-6$~Hz, and a $1/f^{2}$ spectrum at higher frequencies. On the right are exactly the same data but plotted as frequency times power, which better emphasises which frequencies dominate the variance (integrated spectrum). In this $f \times P(f)$ representation a flat spectrum means equal power per logarithmic frequency interval (i.e. $1/f$ spectrum).
This plot also better illustrates the slight deviations from power law form. As is conventional the power density is given in fractional units ([rms/mean]$^2$ Hz$^{-1}$).}
\end{figure}

One of the most fundamental problems in modern astrophysics is to understand the flow of matter accreting onto compact objects such as neutron stars and black holes. These systems are powerful and highly variable sources of X-rays, and it is by studying the fluctuations in X-ray luminosity that we have made progress towards understanding the physics involved. These studies are reviewed by Belloni\cite{belloni} and van der Klis\cite{vanderklis} in this volume. However, almost all this work is dedicated to measuring, modelling and understanding the power spectrum (see Fig.~\ref{fig:psd}) and related second-order statistics (cross spectrum, autocorrelation function, etc.) which necessarily contain information on the statistical moments up to second order only. If the variability processes associated with accretion onto compact objects are linear, Gaussian and stationary then no information is lost (everything is contained in the second order statistics). Until recently this was an implicit and virtually untested assumption, rather than than an open area of investigation, which may in part be due to the fact that constructing physical models of the power spectra has proved to be a serious challenge. However, these assumptions cannot be strictly true since we observe high amplitude fluctuations (rms $\sim 20$ per cent is quite typical) while our knowledge of the physics tells us the X-ray luminosity must be non-negative; the fluctuations in X-ray luminosity cannot be exactly Gaussian if the luminosity is constrained to be always non-negative. Other statistics, that are sensitive to the non-Gaussian and non-linear properties of data, may provide more discriminating tests of existing models or point the way towards new ideas. This article discusses one property of time series from accreting black holes and neutron stars that is not revealed by any individual power spectrum: the rms-flux relation.


\section{THE RMS-FLUX RELATION}
\label{sect:rms-flux}  

\begin{figure}
  \begin{center}
    \begin{tabular}{c}
			\includegraphics[height=5.8cm]{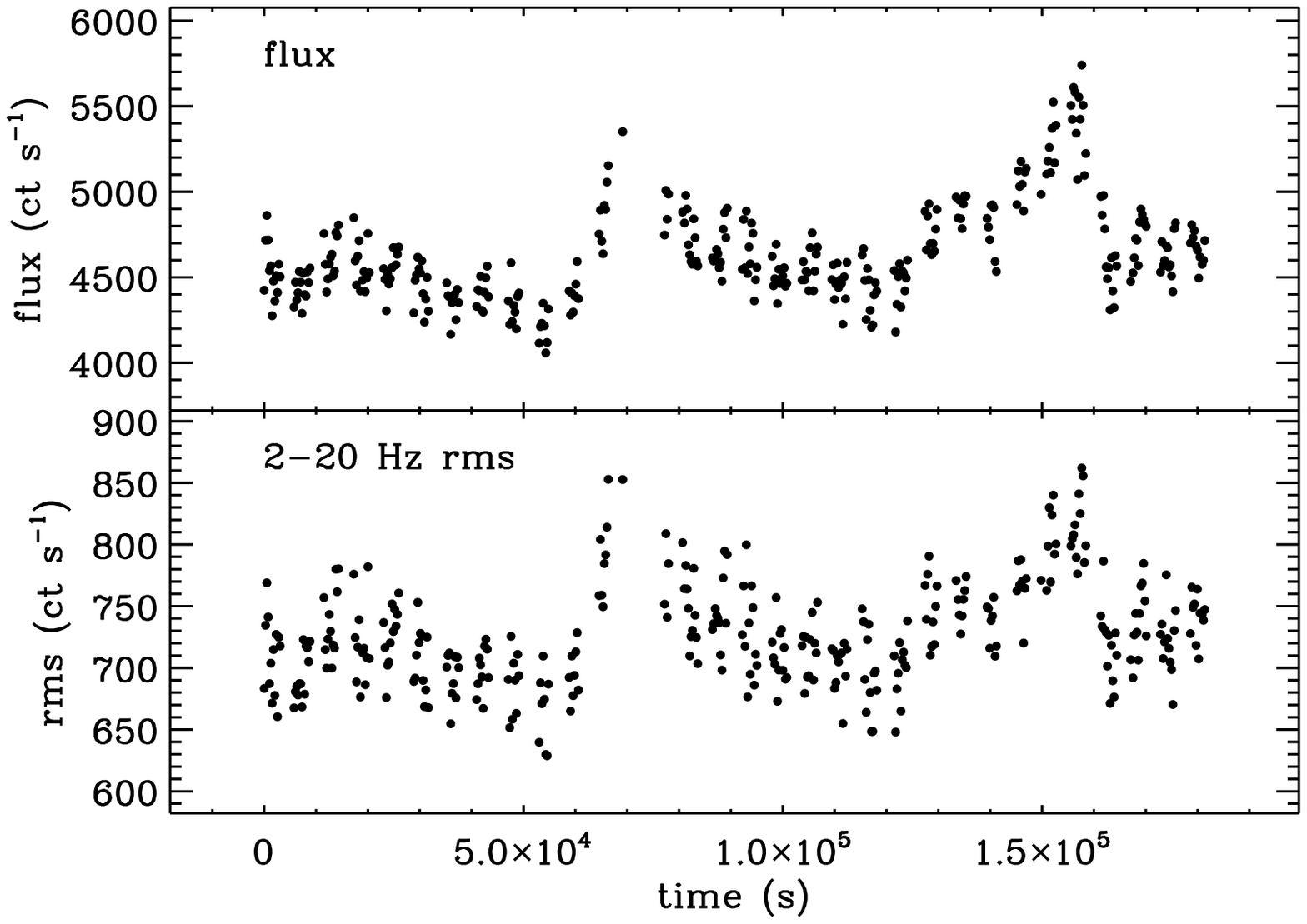}
			\hspace{0.1 cm}
			\includegraphics[height=5.8cm]{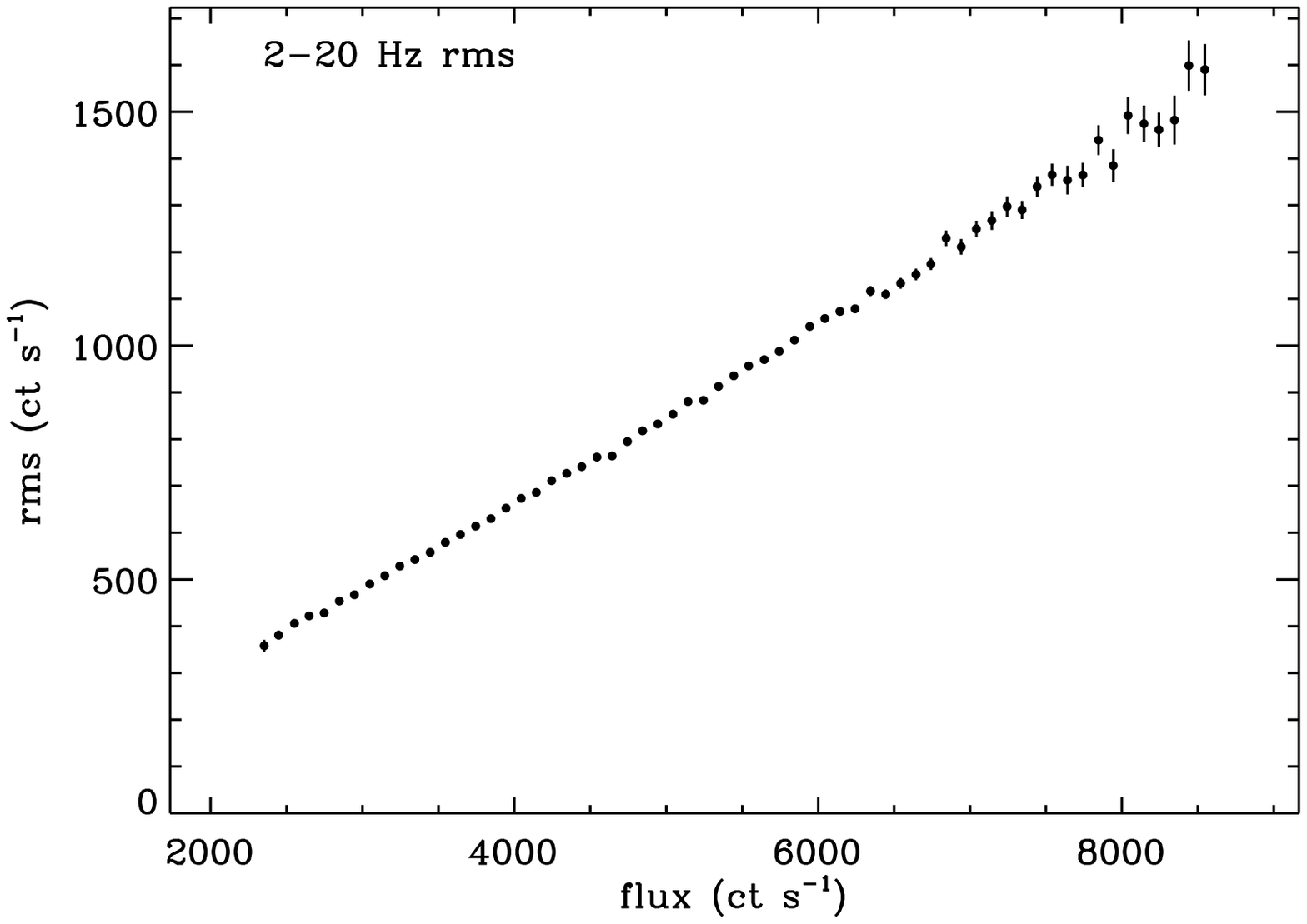}
    \end{tabular}
  \end{center}
  \caption{ \label{fig:rmsflux} 
The rms-flux relation.
The left panels show the time series data from the Dec 1996 observation of Cyg X-1 with \emph{RXTE} which lasted for two days. The gaps in the coverage are due to observational constraints (the satellite is in a low Earth orbit meaning that every $\sim 96$ minute orbit the Earth lies between the target source and the satellite).
The top panel shows the X-ray flux averaged over $256$ s intervals (from data originally binned at $2^{-8}$ sec resolution). The bottom panel shows the $2-20$~Hz rms calculated from the integral of the periodogram of each $256$ s interval. 
The two time series clearly show the same trends, indicating the $2-20$~Hz is correlated with variations in the average flux that occur at much lower frequencies.
However, the two are not identical, of note is the increased scatter in the rms values, which is inherent in values estimated from any finite realisation of a stochastic process.
The right panel shows a plot of the $2-20$~Hz rms averaged in flux bins. In this case both the rms and flux were calculated from non-overlapping $1$ s intervals, and the rms values were averaged in bins of increasing flux. The binning reduces the effect of the scatter in the rms values, and clearly demonstrates the tight linear dependence of the average rms on the flux.
}
\end{figure}

Put simply, the rms-flux relation describes the fact that the root mean square (rms)
amplitude of X-ray variations over some time interval
scales linearly with the average X-ray flux (or brightness) during the same time interval. 
This effect was detected by measuring power spectra for many
short segments of a time series (subtracting off any additional power contributed
by the measurement errors), using these to calculate the variance
over some frequency range $[f_2,f_1]$ by integration (the square root of which is
the rms), and comparing these rms values to the mean flux level for each
segment\footnote{The flux and rms measurements have the same units. In X-ray astronomy 
this is conventionally the detector count rate, i.e. counts s$^{-1}$.}.
Due to the stochastic nature of the data (the power spectra are usually
broad-band noise over several decades in frequency; see
Ref.~\citenum{belloni} and references therein) there is substantial scatter
in the individual rms measurements; averaging the rms measurements in
flux bins reduces this scatter. The result is the average dependence
of rms on `local' flux. When this procedure was applied to data from 
accreting black 
holes and neutron stars taken by the {\it Rossi X-ray Timing Explorer}
({\it RXTE}) a a strong, linear
correlation between rms and flux was revealed\cite{uttley01} (see
Fig.~\ref{fig:rmsflux}).

The linear rms-flux relation has been observed in time series from
stellar-mass black holes and neutron stars\cite{uttley01,gleissner}, 
and subsequently confirmed in active galactic
nuclei\cite{uttley01,edelson,vaughan03a,vaughan03,gaskell,mchardy,vaughan05}. The
linear relation appears to be quite ubiquitous in these systems, 
being present over a range of source `states'\cite{gleissner}, and
also occurring in the highly coherent pulsations from an
accreting millisecond pulsar\cite{uttley04}.
The non-zero offset of the relation (Fig.~\ref{fig:rmsflux}) could
indicate at least two emission components contribute to the observed
emission, one component with the linear dependence between rms and
flux, and another with a constant flux, rms or both.

At first sight this relation may seem trivial, merely a
particularly simple (``pure'') form of heteroscedasticity. 
Indeed, the linearity of the rms-flux relation is equivalent to a form of
scale invariance. If there was no relation between flux and rms (or
a non-linear relation) one
could calculate the flux of a given a time series even if no flux
scale were given. In the absence of a scale parameter one can still
calculate the rms in relative (fractional) units, equivalent to rms/flux, 
which would be a monotonic function of
flux. A given fractional rms would therefore correspond to a unique flux level.
Only in the presence of a linear rms-flux relation, however, is the fractional rms
is independent of flux\footnote{Here we have neglected
any small contribution to the total X-ray emission from a
constant flux or rms component.}, 
and so we cannot assign an absolute scale to the time series
(flux or rms). 

Established tests for heteroscedasticity could have revealed
the correlation and shown it to be
linear\cite{bartlett-kendall,bartlett,box-cox}, but the effect was not
noticed because of another convention in X-ray astronomy.
It is standard practice to plot power spectra in fractional units, i.e. normalise the power
density by the squared flux, which removes the dependence
on the sensitivity of the detector/telescope used (i.e. in relative
units rather than a function of count s$^{-1}$), but this coincidently 
removes the dependence of the power on the flux.

\section{Frequency dependence}
\label{sect:scale}

Before moving on one crucial point needs to be made: the 
rms-flux relation observed from accreting neutron stars and
black holes remains linear on {\it all} timescales it has been tested on, despite
the sometimes complex frequency dependence of the power spectrum.
In particular, there are four frequencies (or timescales) that are important when measuring an
rms-flux correlation. 

Consider a time series $x_k$ of length $N$ sampled at intervals (or averaged over bins)
of length $\Delta t$, such that the total duration of the data is $N \Delta t$.
Now, we can break the time series into consecutive segments of length $n$, giving
$M = N/n$ segments, and for each segment we may calculate the mean level of the
time series $\hat{x}_i$ ($i=1,2,\ldots,M$) and a periodogram $P(f_j) = 2 \Delta t |X_i(f_j)|^2 / n$ (where $X_i$ 
is the Fourier transform of the data in segment $i$, and the factor $2 \Delta t / n$ normalises the periodogram 
such that integrating over positive frequencies gives the variance). By integrating each periodogram
over frequencies $[f_2,f_1]$ we get a sample variance for each segment, $\hat{\sigma}_i^2$, over that frequency range.
For comparing rms and flux there are four frequencies (or timescales) involved:
\begin{itemize}

\item
The duration of the time series $f_4 = 1/N\Delta t$

\item
The length of each segment analysed $f_3 = 1/n\Delta t$

\item
The lower and upper bounds of the integration $f_2, f_1$

\end{itemize}
These are constrained such that $f_4 < f_3 \le f_2 < f_1$.
The constraint that $f_3 \le f_2$ comes about because we cannot integrate a periodogram over
frequencies lower than are measured (i.e. $f_2 \ge 1/n\Delta t$), and similarly $f_1 \le 1/2\Delta t$, which means we cannot integrate over frequencies above the Nyquist limit.
Frequencies $f_4$ and $f_3$ determine range of frequencies covered by the time series of mean values, $\hat{x}_i$, and rms values, $\hat{\sigma}_i$, 
while the frequencies $f_2, f_1$ determine over what range of frequencies the rms is accounting. 

The observed rms-flux relation means that changes in X-ray flux occurring over the frequencies range $[f_4,f_3]$ are correlated with the simultaneous rms measured from the frequency range $[f_2,f_1]$. The relation appears linear irrespective of the values of $f_4,f_3$ and $f_2,f_1$ used.
It is this fact that makes the relation interesting. We could, for example, take a time series from a stationary, linear, Gaussian 
process which has a constant rms, and multiply it by a steep linear function. The effect would be to increase both the mean level and the rms
where the linear function is highest, and decrease both where the linear function is lowest, so as to produce a linear relationship
between rms and (mean) flux. However, on timescales over which the linear function varies by only a small amount, where the variance is
dominated by variations in the original series, there will be no correlation between rms and flux, by construction. In this case the
relation does not exist on all timescales.

The power spectrum for the accreting black hole system Cygnus X-1
is shown in Fig.~\ref{fig:psd}, revealing broad-band noise that has a
red ($1/f$ or steeper) spectrum above $\sim 0.06$~Hz ($\sim 16$-s) and
is roughly white below this (although the power spectrum may steepen
again at substantially lower frequencies). Therefore, time series
segments substantially shorter than $10$ s are red noise, with the
variance (power) dominated by longest timescales (lowest frequencies),
and are weakly non-stationary (the mean can only be well defined on
much longer timescales), even though they are all
realisations of the same stochastic process.
Nevertheless, regardless of the timescales used to measure the mean
`local' flux, and the frequencies used to measure the rms, the two are
linearly related (see Fig.~\ref{fig:psd}). 
As shown by Gleissner et al.\cite{gleissner}, the relation remains
present in data from Cygnus X-1 even when measured on timescales of
months-years. This point is crucial 
for understanding the origin of the relation.

\section{A SIMPLE PHENOMENOLOGICAL MODEL}
\label{sect:simple}  

A linear, Gaussian, stationary, stochastic processes can be constructed by
addition of independent signals in either the time domain (e.g. a
moving average [MA] or shot noise process) or the Fourier domain (by
the sum of Fourier components with independent phases). 
In this scenario there need be no coupling between variations at
different frequencies -- high frequency variations require no
knowledge of the  behaviour of lower frequencies variations. 
This is not true for a red noise process with a linear rms-flux relation on
all timescales, which does require higher frequency variations to
respond to the activity of all lower frequency variations. 
The power spectrum is red noise over most of the available bandpass
and so the mean flux 
in any short time series segment is determined
by the long timescale variations (which dominate the overall
variance), yet the rms on much shorter timescales is correlated with
this. Further demonstrations of this are given in
Refs.~\citenum{uttley01,uttley05}. This suggests a `top-down' process is
required, in which longer timescale (lower frequency) variations
originate first and modulate shorter timescale (higher frequency)
variations. In fact what is required is simple amplitude modulation
whereby the amplitude of variations at any given frequency is
modulated by the variations at all lower frequencies.

This suggests a simple way to synthesise a stochastic time series with the linear
rms-flux relation on all timescales. A general way to synthesise any
linear, Gaussian, stochastic time series is by the summation, in the
time domain, of
individual sinusoidal signals with (independent) random phases. 
\begin{equation}
\label{eqn:sum}
y_k = \sum_{j=1}^{N/2} A_j \sin (2 \pi f_j t_k + \phi_j) 
     = \sum_{j=1}^{N/2} a_j(t_k)
\end{equation}
where $y_k=y(t_k)$ is the output time series and $A_j$ and $\phi_j$
are the amplitude and phase (uniform $\phi_j \in [0,2\pi)$) at frequency $f_j$. 
This is simplified by writing $a_j(t_k) = A_j \sin (2 \pi f_j t_k +
\phi_j)$. For a realistic signal with finite variance $A_j \rightarrow
0$ as $N \rightarrow \infty$, and hence $a_j \rightarrow 0$. In
order to impose a linear rms-flux relation on the output time series
we simply replace the sum with a product. By multiplying the
components (again, in the time domain) we ensure that variations at
any given frequency are modulated by variations on all other frequencies:
\begin{equation}
\label{eqn:product}
x_k = \prod_{j=1}^{N/2} \{ 1 + A_j \sin (2 \pi f_j t_k + \phi_j) \}
\end{equation}
with $0 \le A_j < 1$ to ensure the output signal is positive. 
Consider a simple signal comprising two sinusoids, one low frequency,
one high frequency. The product of these gives an output signal where
the (rms) amplitude on short timescales (dominated by the high frequency
sinusoid) scales with the level of the low frequency sinusoid, which
dominates the mean level, hence produces a linear rms-flux relation.
Extending this procedure we can generate a signal with a 
linear rms-flux relation on all timescales by multiplying sinusoids
from all frequencies (for a power spectrum defined by the $A_i$
values). 

It is interesting to compare the properties of the additive and multiplicative
methods. Taking the logarithm of equation~\ref{eqn:product} we get:
\begin{equation}
\log[x_k] = \sum_{j=1}^{N/2} \log[1+a_j(t_k)] \approx \sum_{i=1}^{N/2} a_j(t_k) = y_k
\end{equation}
where we have used the Taylor series expansion of $\log[1+a] = a - a^2/2 +
a^3/3 + \cdots$ (noting that $|a_j(t_k)| \ll 1$), and kept only the
first order term
(the higher order terms in the Taylor series may be safely neglected,
see Ref.~\citenum{uttley05} for more detail).
This result indicates that taking the logarithm of the signal generated by the
multiplicative method (equation~\ref{eqn:product}) gives the same result as
the additive method (equation~\ref{eqn:sum}). 
Reversing this argument we see that to impose a linear rms-flux
relation on an otherwise Gaussian signal we must only make an
exponential transformation: $x_k \propto \exp[y_k]$.
(This fact could have been obtained from the known variance-stabilising
methods\cite{bartlett-kendall,bartlett,box-cox}, except that it is not
immediately obvious how these methods would be affected by the
non-white power spectrum, i.e. auto-correlation, of the data under consideration here.)

\section{SIMULATIONS OF EXPONENTIALLY TRANSFORMED DATA}
\label{sect:sims}

\begin{figure}[t]
  \begin{center}
    \begin{tabular}{c}
	\includegraphics[height=5.8cm]{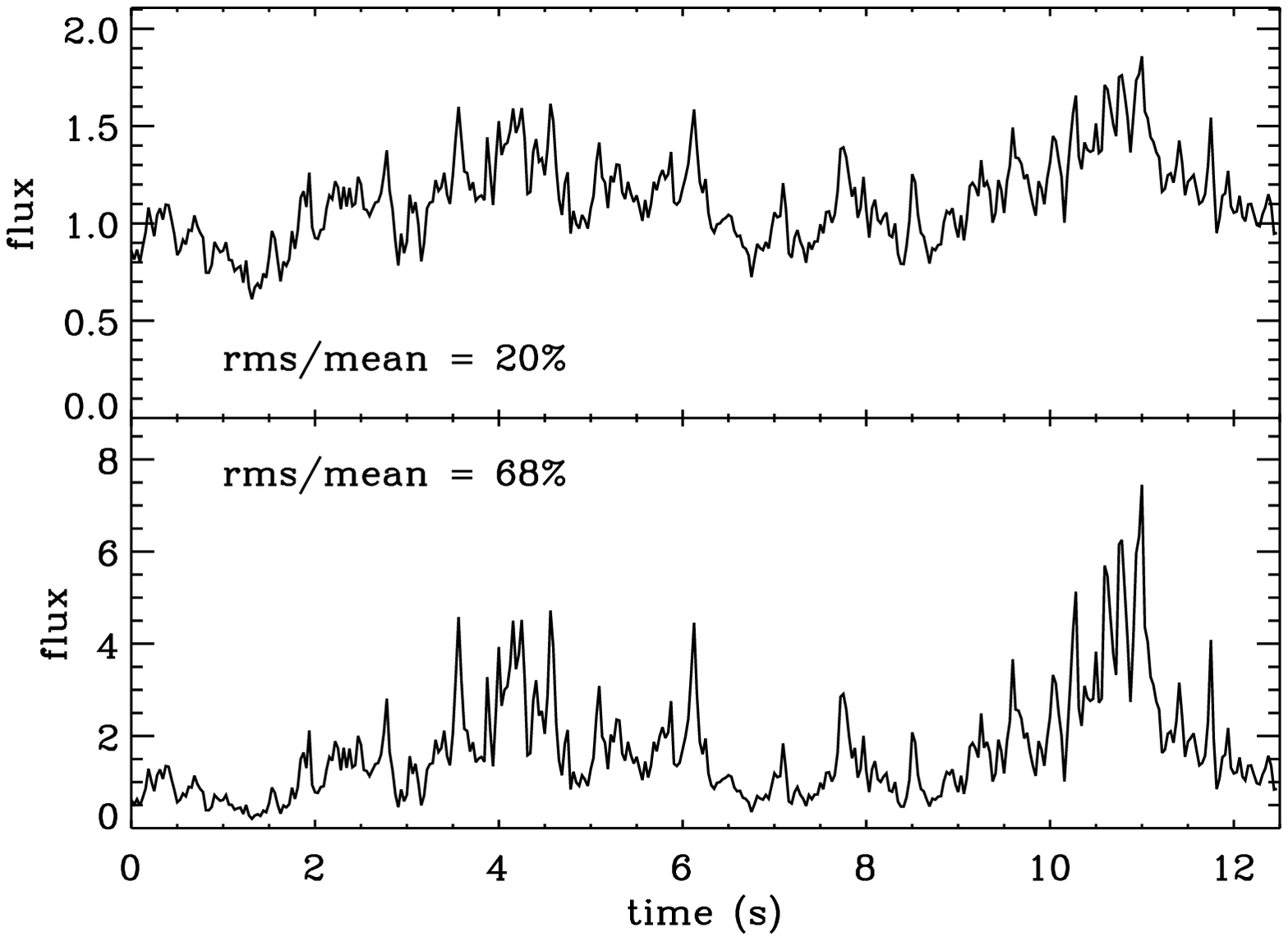}
	\hspace{0.1 cm}
	\includegraphics[height=5.8cm, angle=0]{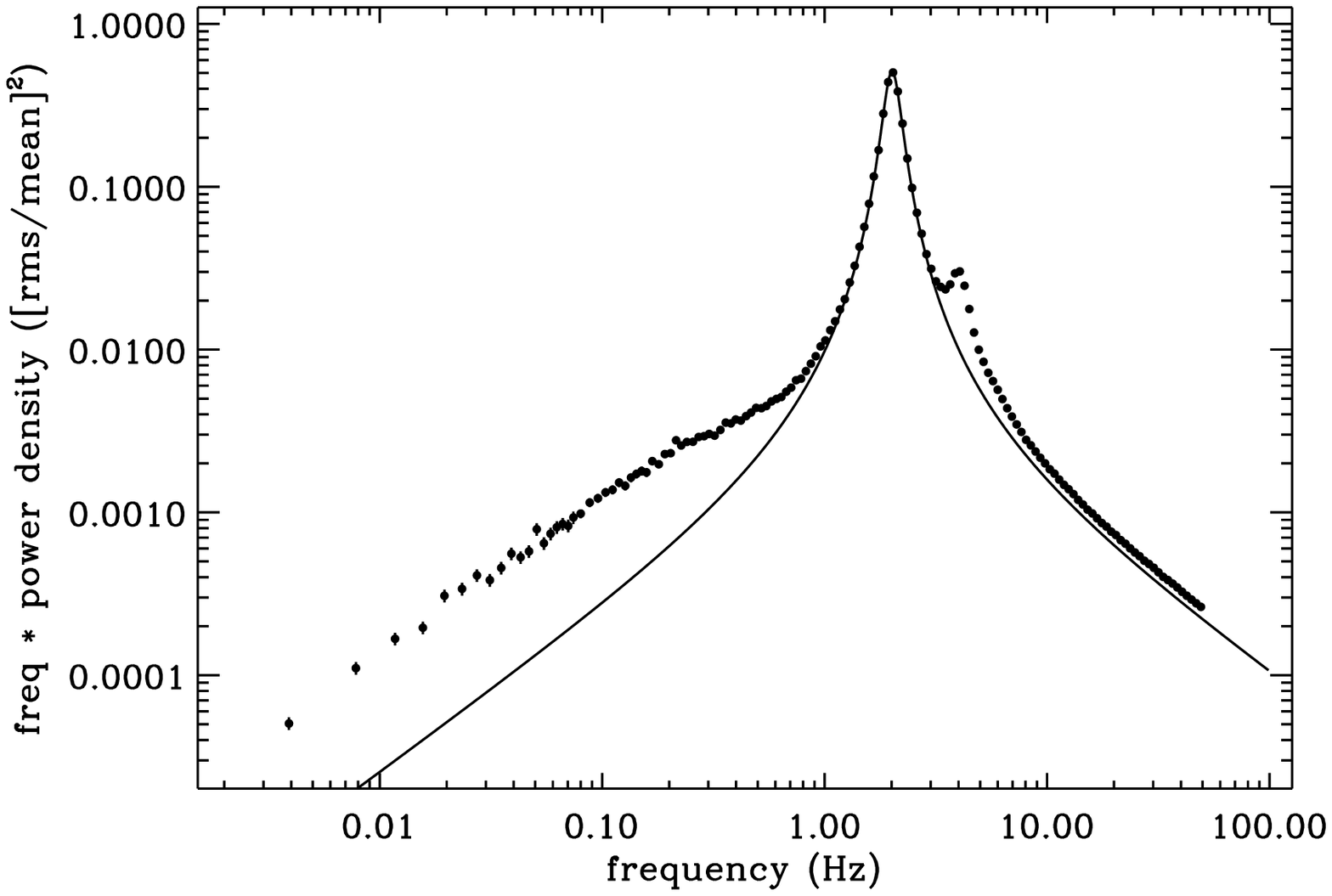}
    \end{tabular}
  \end{center}
  \caption{ \label{fig:sims} 
Exponentially transformed simulated data.
The left panels shows two versions of an artificially generated time series in which the data are the same but the amplitudes (absolute rms levels) were varied prior to exponentially transforming. After applying the exponential transform the two series have different fractional rms values. The data were generated using the power spectrum shown in Fig.~\ref{fig:psd}.
The right panel illustrates the spectral distortion caused by the transformation. The solid curve shows the Lorentzian power spectrum of an input process (linear) and the data show the estimated power spectrum for artificial data after exponential transformation. Note the logarithmic scale that clearly reveals the weak distortions introduced by the transformation, including a harmonic.}
\end{figure}

Now that we have a simple model for generating time series by a simple 
transformation of linear data, we are in a position to produce artificial 
data that obey the rms-flux relation simply by generating linear, Gaussian
data using standard techniques, and the applying the static exponential 
transformation. In this way we can very easily generate data with 
a range of properties (mean, variance, power spectral shape)
and a linear rms-flux relation, then observe the consequences of the relation. 

One of the most widely-used and computationally efficient methods for
generating linear, Gaussian time series with arbitrary power spectrum $P(f)$
is the Davies-Harte method\cite{davies-harte}, where $\sqrt{P(f)}$ 
is used to define the data in the Fourier domain,
and the real and complex components of the Fourier transform are each 
randomised by multiplying by independent normal deviates\footnote{Strictly speaking, 
the method described by Davies \&
Harte\cite{davies-harte} begins by defining the process in term of its 
auto-covariance, then taking the Fourier transform to obtain the power
spectrum $P(f)$. The method of generating linear, Gaussian time series by 
randomising the Fourier transform
and transforming back to the time domain using an FFT was introduced to the astronomical
literature by Timmer \& K\"{o}nig\cite{timmer}.}.
The data are then inverse Fourier transformed to the time domain using an 
FFT. This is a
computationally efficient method for realising the summation of
equation~\ref{eqn:sum} (with random, uniform phases $\phi_j \in [0,2\pi)$ 
and amplitudes distributed as $A_j^2 \sim P(f_j) \chi_2^2 /2$,
where $\chi_2^2$ is a random deviate distributed as chi-square with two degrees
of freedom).

Fig.~\ref{fig:sims} shows short sections of simulated times series produced by 
exponentially transforming data produced with the Davies-Harte method
and a power spectrum taken from Fig.~\ref{fig:psd}. Exactly the same random
number sequence was used to generate the three series, the only difference 
being the total variance (the normalisation of the power spectrum model).
These clearly demonstrate the positively skewed flux distribution
and how this becomes more skewed as the total variance increases. 

The exponential transformation modifies the mean and variance of the data,
therefore we must know the conversion between linear and transformed data
in order to generate transformed data with exactly specified statistical 
properties. The relations between linear and exponential data for the mean 
and variance is the same as those relating the Gaussian and log-normal 
distributions. Specifically, for a linear data $y_k$ with zero-mean ($\mu_y = 0$) 
and variance $\sigma_y^2$, the mean and variance of the exponentially transformed 
$x_k$ data are:
\begin{equation}
\label{eqn:logmean}
\mu_x = \exp[\sigma_y^2 / 2] ~~~ \sigma_x = \exp[\sigma_y] (\exp[\sigma_y] - 1)
\end{equation}

The power spectrum of the transformed data is also not identical to that of the
input linear data (see Fig.~\ref{fig:sims}). For broad-band spectra the
distortion is small, but becomes strongest for strongly peaked spectra
at high amplitudes. The effect of this distortion is described by 
Johnson\cite{johnson} (section III-D), who gives a formula for the effect of transformation
on the auto-covariance function. This can be inverted to give the formula
for the input auto-covariance that produces the desired output auto-covariance in the 
transformed data. 

The above recipe can produce synthetic time series with the linear rms-flux
relation with a wide range of power spectra. However, some processes,
particularly those with very high power concentrated in narrow-band features 
have auto-covariances whose Fourier transforms contain negative values. 
This means they have negative powers, which cannot correspond to a real-valued
time series, meaning no time
series can be generated with the Davies-Harte method. A time series with a strong, narrow-band spectrum
(e.g. a narrow Lorentzian) can be exponentially transformed but the resulting
time series has a spectrum with harmonics. If we try to suppress these
harmonics by using the Johnson formula to specify the spectrum of the transformed 
data as a single, narrow Lorentzian, we find that the recipe fails --
un-distorting the auto-covariance produces a function that has negative-valued
Fourier transform, i.e. negative powers, that cannot be realised in a time 
series. This may indicate that it is not possible for Nature to produce a 
strong, narrow band signal without additional harmonics if the linear rms-flux relation
is present (Fig.~\ref{fig:sims}).

\section{CONSEQUENCES OF THE EXPONENTIAL TRANSFORMATION}
\label{sect:predict}  

If we consider the linear rms-flux relation in purely signal
processing terms, it can be seen as an effect of the exponential
transformation of a linear, Gaussian signal (e.g. the output of
equation~\ref{eqn:sum}). This has several consequences. 

\begin{figure}
  \begin{center}
    \begin{tabular}{c}
	\includegraphics[height=5.8 cm]{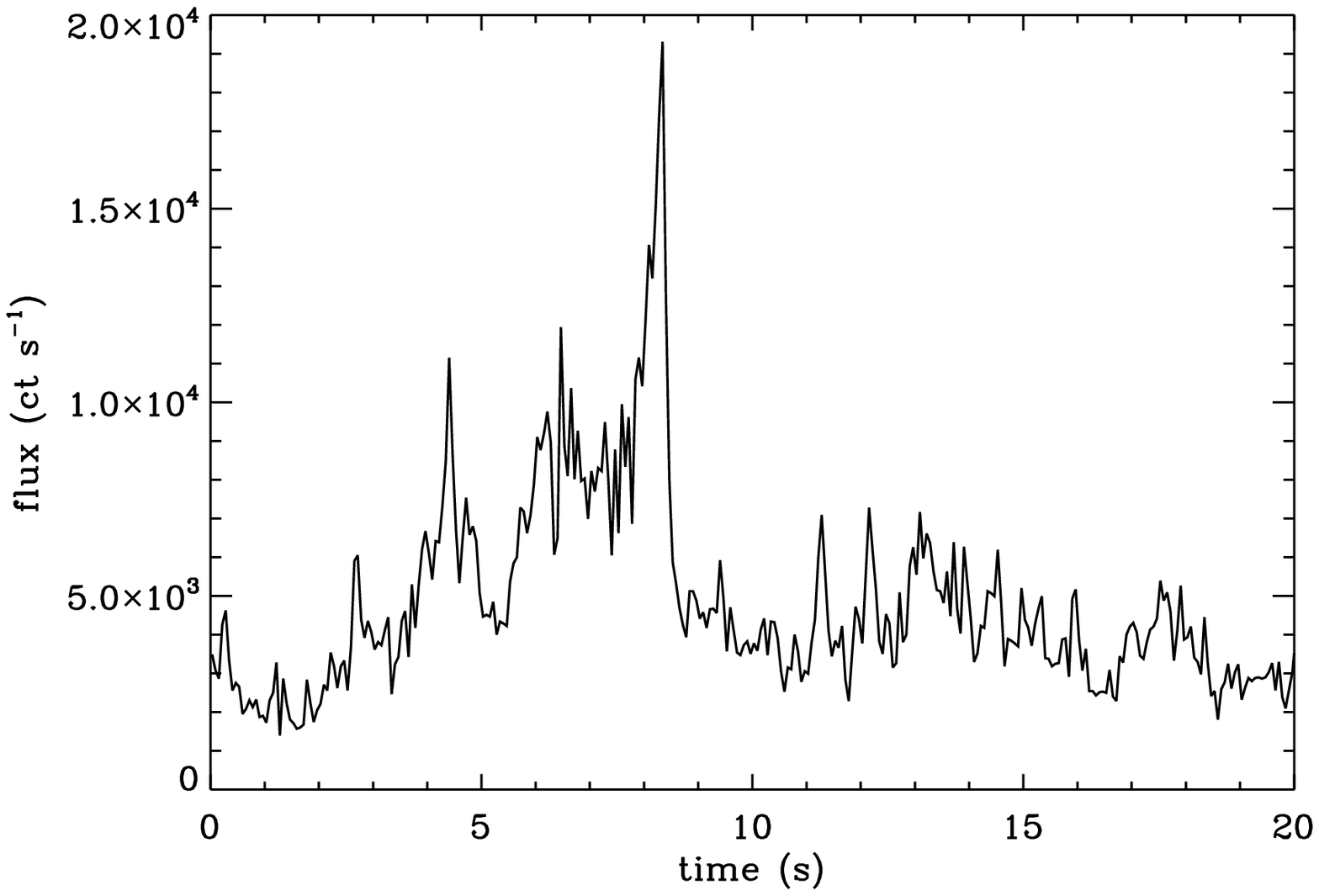}
	\hspace{0.1 cm}
	\includegraphics[height=5.8 cm]{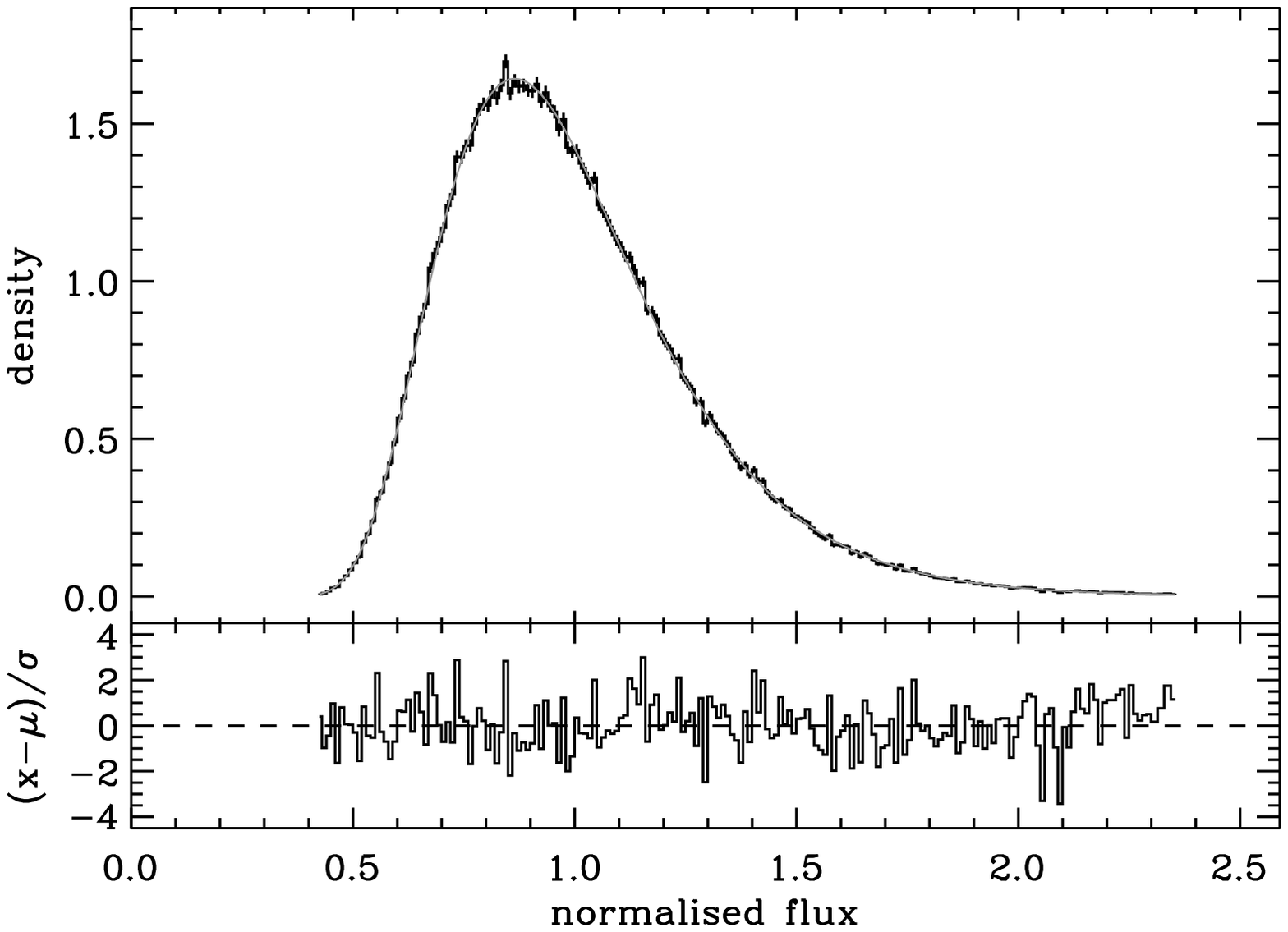}
    \end{tabular}
  \end{center}
  \caption{ \label{fig:lognormal} 
The left panel shows an X-ray time series from Cygnus X-1. Specifically, this is a $20$ s interval from the start of the much longer Dec 1996 observation, and for clarity the data were rebinned on a scale of $0.0625$ s ($2^{-4}$~s) from data originally sampled at $2^{-8}$ s. Compare with the simulations of Fig.~\ref{fig:sims}.
The right panel shows the empirical density function for X-ray fluxes for Cygnus X-1 from December 1996. The histogram (with error bars) shows the empirical density and the solid grey curve is the best-fitting lognormal distribution, and the bottom panel shows the normalised data-model residuals from this fit. The lognormal fit to the histogram gave $\chi^2 = 238.6$ for $189$ degrees of freedom, not a great fit, but good considering the very high signal-to-noise.
The empirical density is the normalised histogram of fluxes averaged over $0.25$ s long time intervals. 
For each continuous segment of data ($\sim 3000$ s per \emph{RXTE} orbit) the fluxes at $0.25$ s intervals were normalised to the mean level for that particular data segment --  this reduces any slight shift in the mean of the distribution due to very low frequency power (or non-stationary trends). This gave a total of $358,524$ flux samples that were used to produce the histogram. }
\end{figure}

{\bf Log-normality:~}
The linear rms-flux relation is equivalent to an exponential transformation 
of Gaussian data, which in turn
predicts a log-normal distribution of fluxes (over long
timescales), since the exponential transformation of a Gaussian
distribution is log-normal. The long observation of Cygnus X-1 taken in
December 1996 provides a stunning confirmation of this prediction
(Fig.~\ref{fig:lognormal}; Ref.~\citenum{uttley05}). However, one must again be careful to
account for the weak non-stationarity of signals with red power
spectra, for which the flux distribution is dominated by the lowest
frequency variations, which are not sufficiently sampled to converge
on the correct underlying distribution. Only by measuring the
distribution over sufficiently long timescales that the power spectrum
is nearly flat (white) can we be sure to completely sample the true
distribution. 

{\bf Non-linearity:~}
Another consequence of the exponential transformation model is that
time series showing a linear rms-flux relation on all timescales are,
in one sense, non-linear\footnote{
Here we should note that only models for processes can be 
non-linear, and not time series data themselves, strictly speaking.
What we really mean is that this simple phenomenological model that
reproduces the main features of these time series (power spectrum
shape and linear rms-flux relation) is non-linear.}. 
Any linear, stochastic model may be
written as an MA. In discrete notation:
\begin{equation}
\label{eqn:ma}
y_k = \sum_{i=0}^{\infty} g_i \varepsilon_{k-i}
\end{equation}
where $\varepsilon_k$ are independent and identically distributed
({\it iid}) random deviates, and $g_i$ is the kernel that defines the
`memory' of the process (see Ref.~\citenum{priestley} for details). A
non-linear model cannot be written in this form but requires
higher order terms. With some generality, a non-linear model may be
written as a Volterra series\cite{priestley}:
\begin{equation}
\label{eqn:volterra}
x_k = \sum_{a=0}^{\infty} G_a \varepsilon_{k-a} 
    +  \sum_{a=0}^{\infty} \sum_{b=0}^{\infty} G_{ab} \varepsilon_{k-a} \varepsilon_{k-b}
    +  \sum_{a=0}^{\infty} \sum_{b=0}^{\infty} \sum_{c=0}^{\infty}  
           G_{abc} \varepsilon_{k-a} \varepsilon_{k-b}\varepsilon_{k-c}
    + \cdots
\end{equation}
The model is linear if the terms $G_{ab}, G_{abc}, \ldots$ are all
zero (in which case it reduces to equation~\ref{eqn:ma}). This is
not true if we consider the exponential transformation of a general
linear process:
\begin{equation}
x_k = \exp[y_k] = 1 + y_k + \frac{(y_k)^2}{2} + \sum_{n=3}^{\infty} \frac{(y_k)^n}{n!}
   = 1 + \sum_{i=0}^{\infty} g_i \varepsilon_{k-i} +
      \frac{1}{2} (\sum_{i=0}^{\infty} g_i \varepsilon_{k-i})^2 +
      \sum_{n=3}^{\infty} \frac{1}{!n}  (\sum_{i=0}^{\infty} g_i \varepsilon_{k-i})^n
\end{equation}
The last equality made use of the expression of a
general linear model (equation~\ref{eqn:ma}). 
This may be re-arranged to form a Volterra series:
\begin{equation}
x_k = 1 + \sum_{a=0}^{\infty} G_a \varepsilon_{k-a} 
    +  \sum_{a=0}^{\infty} \sum_{b=0}^{\infty} G_{ab} \varepsilon_{k-a} \varepsilon_{k-b}
    +  \sum_{a=0}^{\infty} \sum_{b=0}^{\infty} \sum_{c=0}^{\infty}  
           G_{abc} \varepsilon_{k-a} \varepsilon_{k-b}\varepsilon_{k-c}
    + \cdots
\end{equation}
with the coefficients $G_{a} \propto g_{a}$, $G_{ab} \propto g_{a}g_{b}$, $G_{abc} \propto g_{a}g_{b}g_{c}$, and so on. These higher order terms are non-zero if $g_i$ is non-zero for more than one $i$ (i.e. the power spectrum is not white), meaning the exponential transform of a general linear process cannot (in general) be written in terms of a linear model. This argument is discussed more more detail by Ref.~\citenum{uttley05}.

The fact that accreting compact objects do show such a linear rms-flux relation on all observable timescales may mean that their time series can be modelled in terms of a linear, Gaussian process subject only to an exponential transformation, which after all is a static non-linear transformation, and suggests there is no need for dynamical non-linearities. 

{\bf Millisecond flares:~}
Cygnus X-1, and perhaps other X-ray binaries, occasionally display strong flares, bursts of increased X-ray activity lasting $\sim 0.1$ seconds. These have been identified as being inconsistent with the variabilityarising from a stationary, Gaussian process\cite{gierlinski}, but have been shown to be consistent with the output of an exponentially transformed Gaussian process\cite{uttley05}.

{\bf Prolonged `low flux' episodes in AGN:~}
Some of the most well-studied Active Galactic Nuclei 
(AGN; see Ref.~\citenum{uttley07} in this volume) display 
episodes of very low activity lasting for days-months, when the X-ray
flux remains low and relatively stable\cite{uttley99}. These can be easily understood
as a consequence of the linear rms-flux relation. At high fluxes the
variance is also high, and so the source appears very X-ray `active',
while at much lower fluxes the variance also also at a minimum, 
making the source appear to be in a relatively stable `low' state 
(see Fig.~\ref{fig:sims}). The difference between `high' and
`low' activity periods becomes more exaggerated as the gradient of
the rms-flux relation increases. This gradient is equivalent to 
the fractional rms - those sources with strongest overall
variability are also those with the most marked transitions from
`active' to `inactive' periods (on an absolute scale).

\section{PHYSICALLY MOTIVATED MODELS}
\label{sect:physical}  

The presence of a linear rms-flux relation on all timescales has some important
consequences for our understanding of the physics involved in
producing variable X-ray emission from accreting compact objects.
The relation appears the same in all known accreting black hole systems
(stellar mass\cite{belloni} and supermassive\cite{uttley07} black holes),
and also in at least one accreting neutron star system\cite{uttley04}. As such
it seems rather general to accreting compact objects, and any physical model
of the variability mechanism must reproduce not only the correct power
spectrum but also the rms-flux relation in order to be acceptable. One important
point that can be made immediately is that models involving dynamical non-linearities
(chaos) are not required: The non-linear behaviour of accreting systems can be
explained by a simple, static non-linear transformation, eliminating the need for
dynamical non-linearities.

\subsection{Models that fail the rms-flux test}

Almost since the discovery of randomly fluctuating X-ray brightness
from Cygnus X-1 the most popular models for the X-ray
variability have all been variants of shot noise\cite{terrell}. By construction,
shot noise can reproduce any power spectral shape, in the same way that 
an MA process can have any conceivable power spectrum. Therefore 
a shot noise model is always capable
of matching the power spectra of time series from accreting compact
objects, and it is quite straightforward to imagine individual and
independently occurring shots as the result of `active regions' appearing
and disappearing around or on an accretion disc in a manner not dissimilar to
X-ray emitting flares above the surface of the Sun. 

But while shot noise can reproduce any power spectrum, it also 
predicts a roughly Poisson or Gaussian distribution of X-ray fluxes. This is because the
observed time series do not appear as a series of clearly
resolved events or shots but as a fluctuating continuum, as is
reflected in the very broad power spectrum.
In the limit of a large number of independent shots occurring at any one time
the output time series will be linear and Gaussian (by the central
limit theorem) and will not show any dependence of rms on flux nor a lognormal flux distribution.

A closely related model is Self-Organised Criticality\cite{bak87,bak88,christensen} (SOC),
where certain systems of interacting elements reach a critical point at which
events occur on all possible scales with a power law distribution. 
Indeed, there have been attempts to model X-ray emission from accretion discs with
toy models based on SOC\cite{mineshige,takeuchi}. But, as pointed out by 
Bak et al.\cite{bak88} (see also section 4 of Christensen et al.\cite{christensen}),
SOC produces power law distributions of event scales.
These events need to be randomly superposed to form shot noise with a broad
power spectrum (e.g. $\sim 1/f$).
By the central limit theorem, the resulting flux distribution will be Gaussian again.
SOC is a model for generating a power law distribution of event scales, not a model
for interaction of events.

\subsection{Models that pass the rms-flux test}

As described above, a model involving the modulation (multiplication) of
independent fluctuations, rather than the superposition (addition), can
reproduce the linear rms-flux relation. One model of accretion flows that
has this property was discussed by Lyubarskii\cite{lyubarskii}. In this model
random fluctuations in the accretion rate occur at all radii and propagate
through the disc. 
The X-ray emission is produced only from the inner edge of the disc,
near the compact object (black hole or neutron star),
and responds in proportion to the total mass accretion rate at the
inner edge, which is the product of the fluctuations propagating in
from all radii.
Longer timescale (lower frequency) fluctuations 
are generated in the outer regions and propagate inwards modulating the 
more rapid fluctuations generated in inner regions. 
This model has been developed further\cite{churazov,kotov} and 
can reproduce many of the observed timing properties including the
linear rms-flux relation. Another model that also produces amplitude
modulated fluctuations was discussed by King et al.\cite{king}.

\begin{figure}
  \begin{center}
    \begin{tabular}{c}
	\includegraphics[height=5.8cm]{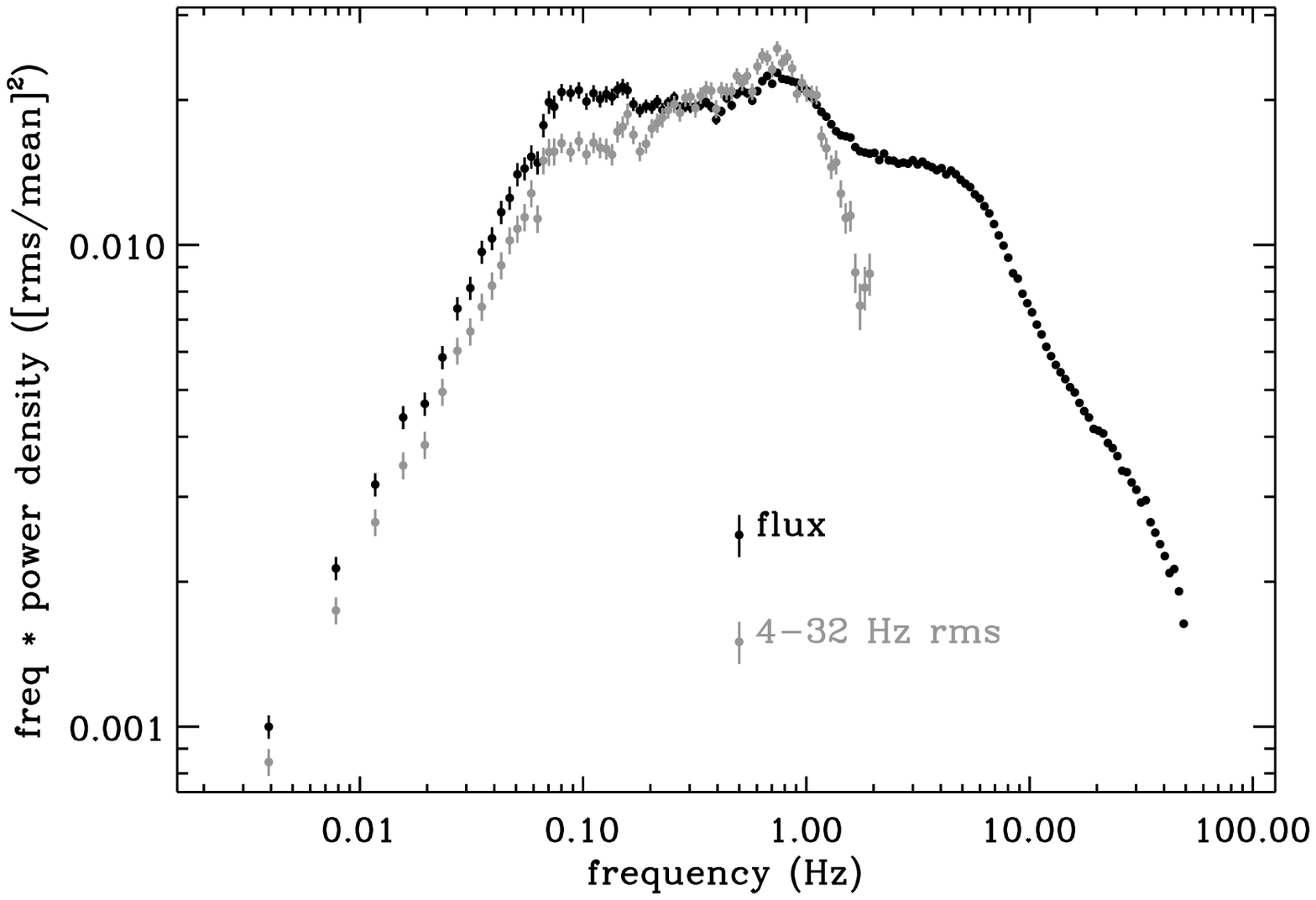}
	\includegraphics[height=5.8cm]{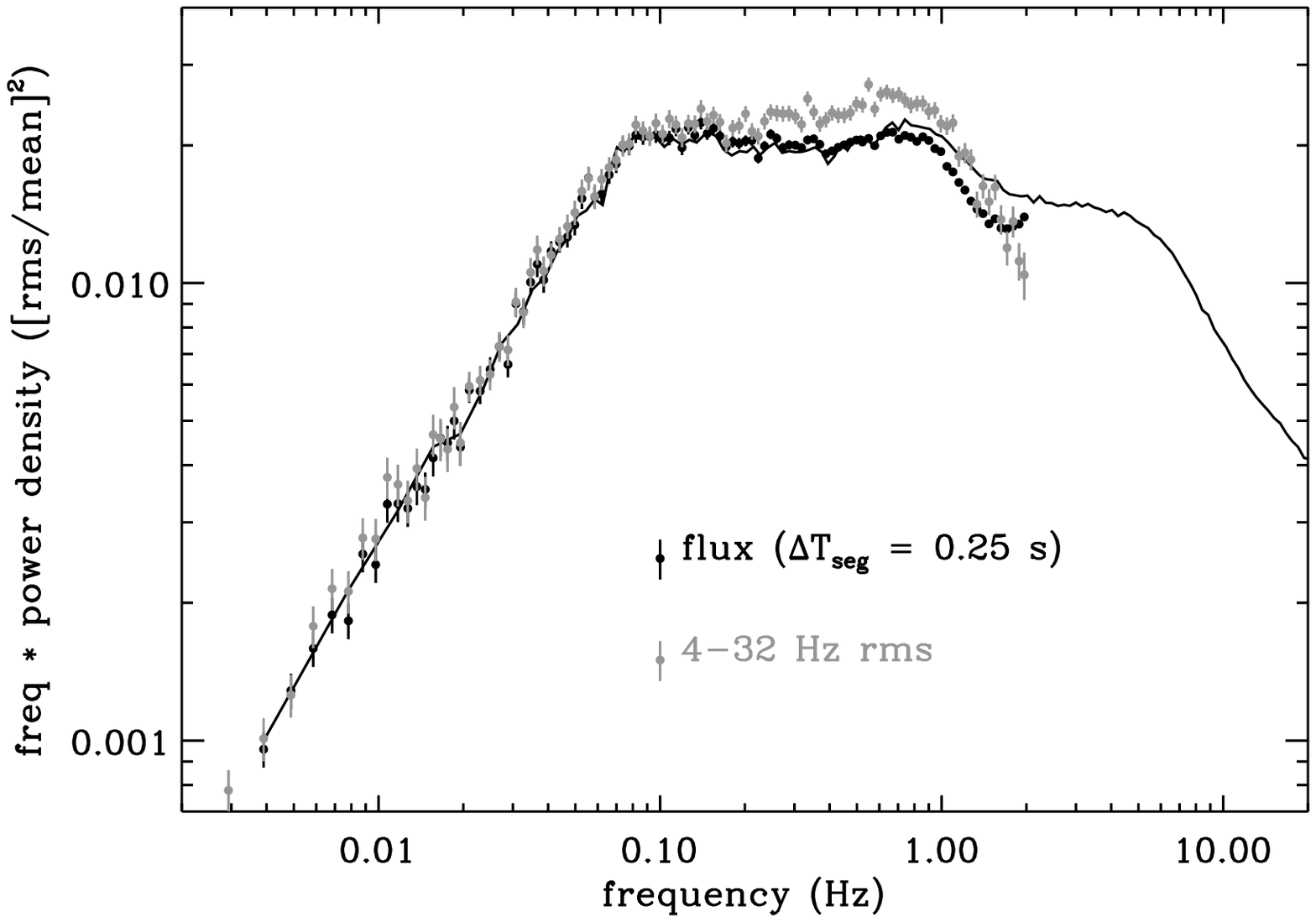}
    \end{tabular}
  \end{center}
  \caption{ \label{fig:rmspsd} 
The right panel shows power spectra for two time series derived from the same Cygnus X-1 data: (black)  the time series of X-ray fluxes (as Figure.~\ref{fig:psd}) and (grey) a time series of $4-32$~Hz rms amplitudes. Both were calculated using the same $0.25$ second segments. Clearly the spectrum of the rms is very
similar to the spectrum of the flux, except for at the highest frequencies. (The estimated spectra from both the flux and rms data contain flat `background' noise terms that must be subtracted to reveal the spectrum of the true process. In the former case the background noise is due to Poisson noise in the collection of X-ray counts by the detector, in the latter case it is due to random variations in the variance inherent in any finite realisation of a stochastic process.)
The right panel shows the results of the same analysis applied to artificial data. The solid curve shows the power spectrum used to produce linear, Gaussian data (from the measured power spectrum of the real Cyg X-1 data). Time series generated from this spectrum were exponentially transformed, and the mean flux and rms measured from segments of length $0.25$ s, exactly as for the real data. The effect of the transformation, and also the averaging, slightly distorts the spectrum of the mean fluxes away from the original input spectrum. But it is clear the power spectrum of the rms is similar between the real data and the transformed artificial data, indicating that the exponential transformation is largely (but perhaps not entirely) responsible for the structure in the spectrum of the rms.
}
\end{figure} 

\section{OTHER DIAGNOSTICS}
\label{sect:other}  

If the rms of a signal, calculated over any given frequency range, is linearly correlated with the signal's strength, then the rms values should form a time series with the same power spectrum as that of the signal itself. This can be tested by estimating the power spectrum of the rms time series. There is a complication due to the scatter in rms values caused by the stochastic nature of the underlying process. However, if the power spectrum of the process is known (or can be estimated accurately, as is the case here), then the amplitude of these random fluctuations on the rms can be calculated, and its effect of the power spectrum of the rms can be subtracted. Fig.~\ref{fig:rmspsd} shows the power spectrum of the signal (source flux) compared to the power spectrum of the rms (computed over $4-32$ Hz). Clearly the two are very close, but not identical. The slight difference in normalisation at lower frequencies is mostly due to the presence of Poisson noise from the detection process, which adds an approximately constant amount to the rms (and therefore lowers the fractional amplitude of the frequency dependent variations). The divergence at higher frequencies ($> 1$ Hz) is more intriguing. The right panel shows the results of the same procedure applied to data simulated using the observed power spectrum, and transformed using the exponential transform. As predicted the power spectrum for the signal and rms match closely. The difference between the real data and simulated data may indicate the coupling between flux and rms is not quite as simple as predicted by the exponential transformation. (Another technique not yet applied to X-ray data is to calculate the coherence between two time series, one for the mean flux and one for the rms, to better understand their coupling.)

Another potentially powerful diagnostic of frequency coupling is the bispectrum, or bicoherence.
Where the power spectrum is a second-order spectrum, formed from the product of two Fourier transforms [$X(f_j)$ $X^{\ast}(f_j)$], the bispectrum is a third-order spectrum formed from the product of three Fourier transforms [$X(f_j)$ $X(f_k)$ $X^{\ast}(f_j + f_k)$]. The computation and interpretation of bispectra (and higher-order spectra in general) is treated by Refs~\citenum{brillinger,nikias}, and they have been applied to data from various fields including geology\cite{rial}, plasma physics\cite{kim} and speech processing\cite{fackrell}.

The bispectrum is a complex quantity defined for the pair of frequencies $(f_j,f_k)$.  The phase of the bispectrum is often called the \emph{biphase} and its modulus-squared amplitude is called the \emph{bicoherence} (once it has been suitably normalised).  The bispectrum provides a measure of the multiplicative nonlinear interaction of frequency components in the time series\cite{hinich}.  A simple way to understand the bicoherence is as follows.  If the variances occurring at the frequencies $f_j, f_k, f_j+f_k$ are independent (spontaneously excited fluctuations) they will have independent, random phases. The bispectrum will thus tend to zero.  If there are variations occurring at $f_j+f_k$ that are due to multiplicative coupling of variations at $f_j$ and $f_k$ then there will be phase coupling and the bispectrum will be non-zero. The value of the bicoherence in some sense represents the faction of variability  power due to multiplicative coupling. A stationary, linear, Gaussian processes has a bispectrum that vanishes (at all frequencies in the large sample limit), and therefore the amplitude of the bicoherence is (asymptotically) zero. For stationary, linear, non-Gaussian process  bicoherence is a non-zero constant while for a stationary
non-linear process the bicoherence is not constant with frequency. These properties make the bispectrum sensitive to the non-Gaussian and non-linear characteristics of the time series\cite{hinich}. 

\begin{figure}
  \begin{center}
    \begin{tabular}{c}
	\includegraphics[height=5.8cm]{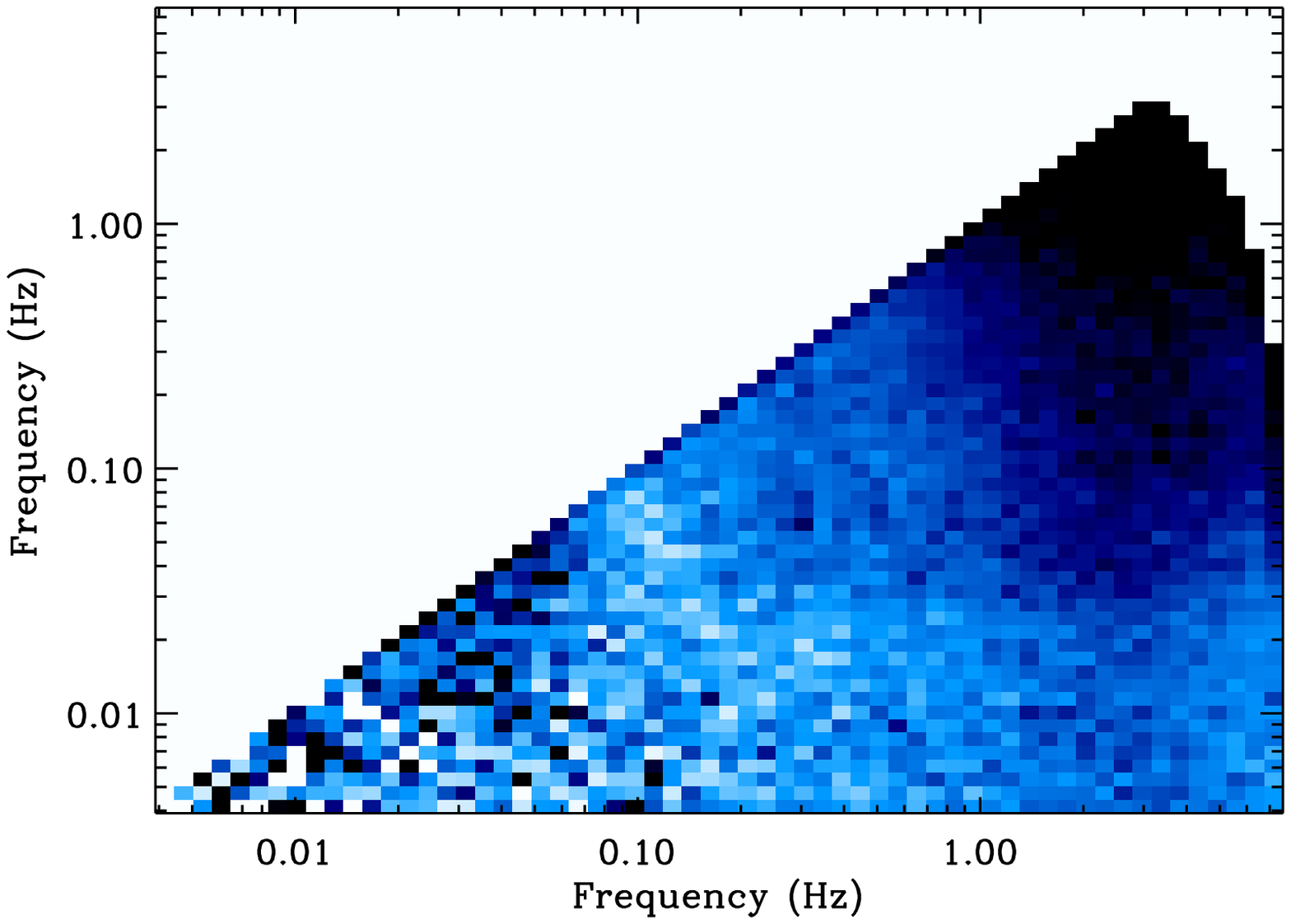}
	\includegraphics[height=5.8cm]{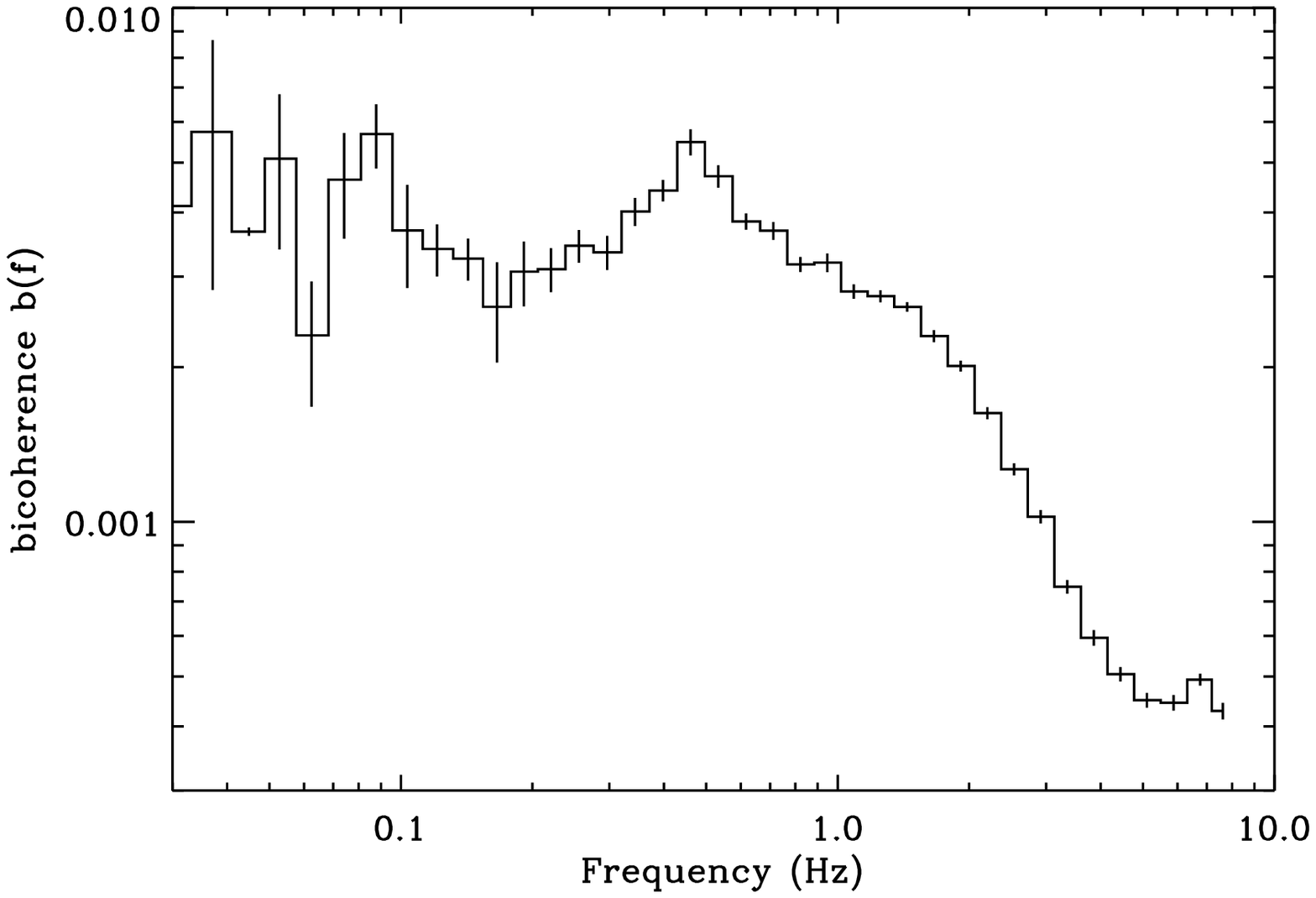}
    \end{tabular}
  \end{center}
  \caption{ \label{fig:bicoh} 
Bicoherence for the Cygnus X-1 data. The left panel shows the biboherence from the `principal domain' of bi-frequencies $f_j,f_k$ (the principal domain is defined as those pairs of frequencies for which $f_j$ spans the full range of Fourier frequencies $[0,1/2\Delta t]$, with the additional constraints $f_k \le f_j$ and $f_j + f_k \le 1/2\Delta t$). The effect of detector noise has been subtracted (to first order). The bicoherences were binned in equal logarithmic frequency intervals along each axis, and colour coded such that lighter shades correspond to higher bicoherences. In order to better illustrate the structure in the bicoherence the right panel shows the bicoherence compressed into one dimension as a function of $f_j + f_k$ (by summing over lines of constant $f_j+f_k$). The error bars come from the scatter of individual bicoherences within each logarithmic frequency bin, but as individual bicoherence estimates are not independent at different bifrequencies, these should be taken as illustrative only.
}
\end{figure}

The success of the amplitude modulation description of the Cygnus X-1 data suggests that fluctuations at different frequencies are coupled, meaning the rms-flux relation should accompany structure in the bicoherence. Fig.~\ref{fig:bicoh} clearly reveals this to be true;  the non-trivial structure in the bicoherence indicates non-linear frequency coupling. However, interpreting the bicoherence is not simple. The bicoherences at each bifrequency $( f_j, f_k )$ are not independently distributed, making it difficult to quantify error bars and therefore apply model-fitting procedures. Furthermore, the presence of Poisson noise in these data causes spurious bicoherence at high frequencies because while the noise is uncorrelated with the signal (i.e. the true source flux), it is not independent, since the amplitude of Poisson fluctuations depends on $\sim \sqrt{flux}$.
At high frequencies the Poisson noise dominates over the intrinsic variations in the source flux (which drop off rapidly with increasing frequency).
Nevertheless, bicoherence may become a powerful test for more advanced models of variability from black holes\cite{maccarone,uttley05,maccarone05}.

\section{ANALOGUES IN NATURE}
\label{sect:nature}  

Apart from accreting compact objects, do other natural processes that have broad band noise spectra and also show 
a linear rms-flux relation, or is this something peculiar to the dynamics of accretion flows?

Zhang\cite{zhang} has made comparisons with Solar flares and the flaring behaviour of Gamma-Ray Bursts (GRBs), but these are transient events and do not have well-defined broad band noise spectra. As a result of this, it is not possible to conclude whether a linear rms-flux relation occurs on all timescales (fluctuations at all frequencies modulate each other) or whether it occurs because there is an overlying exponential decay that modulates small scale flickering. For example, we could produce a time series using linear shot noise and then multiply this by a strong exponentially decaying trend. The trend dominates the average level and also modulates the variance from fluctuations at higher frequencies. But beyond this, there is no further modulation by fluctuations of different frequencies. Exactly because of this many economic time series, such as the historical value of the S\&P 500 share index, appear at first sight to display the linear rms-flux relation. However, this is the result of inflation -- a dominating, long-term increase in the scale. It is not clear whether random fluctuations on shorter timescales modulate random fluctuations on yet shorter timescales.

Some natural phenomena that are known to possess a strong coupling between signal strength and rms.
Argollo de Menezes and Barab\'{a}si\cite{demenezes} discuss how various the time-dependent traffic on internet routers of the Mid-Atlantic Crossroads network, and stream-flow on rivers  in the U.S. river basin, both obey a relation such that the rms of the activity of a sub-system (router or river) scales with the mean level of activity of that subsystem. The scaling is a power law, but in some cases the index is unity, giving a linear rms-flux relation. This is similar to ``Taylor's law" in population dynamics\cite{Taylor} and is a well known property of many ecological\cite{mcardle} and financial systems. 
The difference between these results and those for accreting X-ray sources is that in the case of the networks, the relation describes how mean and rms values over different subsystems (be they species, routers or rivers) are related, whereas for the X-ray sources we are looking a mean and rms values at different times for just one system. (Although the same relation occurs in many systems.)
Despite this difference it may still be possible to compare the characteristics and interpretation of rms-flux relations in these very different systems.

\acknowledgments     
 
The authors wish to acknowledge financial support through the UK
Science and Technology Facilities Council (STFC).


\bibliography{report}   
\bibliographystyle{spiebib}   

\end{document}